\documentclass[aps,prd,twocolumn,showpacs,nofootinbib]{revtex4-1}

\usepackage{amssymb} \usepackage{amsmath} \usepackage{graphicx}
\usepackage{epsfig,latexsym}
\usepackage{slashed}
\usepackage{mathtools}% http://ctan.org/pkg/mathtools
\RequirePackage{xspace} \allowdisplaybreaks

\begin{document}

\def\bef{\begin{figure}}
\def\eef{\end{figure}}

\newcommand{\nl}{\nonumber\\}

\newcommand{\ans}{ansatz }
\newcommand{\be}[1]{\begin{equation}\label{#1}}
\newcommand{\beq}{\begin{equation}}
\newcommand{\ee}{\end{equation}}
\newcommand{\beqn}[1]{\begin{eqnarray}\label{#1}}
\newcommand{\eeqn}{\end{eqnarray}}
\newcommand{\bd}{\begin{displaymath}}
\newcommand{\ed}{\end{displaymath}}
\newcommand{\mat}[4]{\left(\begin{array}{cc}{#1}&{#2}\\{#3}&{#4}
\end{array}\right)}
\newcommand{\matr}[9]{\left(\begin{array}{ccc}{#1}&{#2}&{#3}\\
{#4}&{#5}&{#6}\\{#7}&{#8}&{#9}\end{array}\right)}
\newcommand{\matrr}[6]{\left(\begin{array}{cc}{#1}&{#2}\\
{#3}&{#4}\\{#5}&{#6}\end{array}\right)}
\newcommand{\cvb}[3]{#1^{#2}_{#3}}
\def\lsim{\raise0.3ex\hbox{$\;<$\kern-0.75em\raise-1.1ex
e\hbox{$\sim\;$}}}
\def\gsim{\raise0.3ex\hbox{$\;>$\kern-0.75em\raise-1.1ex
\hbox{$\sim\;$}}}
\def\abs#1{\left| #1\right|}
\def\simlt{\mathrel{\lower2.5pt\vbox{\lineskip=0pt\baselineskip=0pt
           \hbox{$<$}\hbox{$\sim$}}}}
\def\simgt{\mathrel{\lower2.5pt\vbox{\lineskip=0pt\baselineskip=0pt
           \hbox{$>$}\hbox{$\sim$}}}}
\def\unity{{\hbox{1\kern-.8mm l}}}
\newcommand{\eps}{\varepsilon}
\def\ep{\epsilon}
\def\ga{\gamma}
\def\Ga{\Gamma}
\def\om{\omega}
\def\omp{{\omega^\prime}}
\def\Om{\Omega}
\def\la{\lambda}
\def\La{\Lambda}
\def\al{\alpha}
\newcommand{\ov}{\overline}
\renewcommand{\to}{\rightarrow}
\renewcommand{\vec}[1]{\mathbf{#1}}
\newcommand{\vect}[1]{\mbox{\boldmath$#1$}}
\def\tm{{\widetilde{m}}}
\def\mcirc{{\stackrel{o}{m}}}
\newcommand{\Dm}{\Delta m}
\newcommand{\dm}{\varepsilon}
\newcommand{\tanb}{\tan\beta}
\newcommand{\nbar}{\tilde{n}}
\newcommand\PM[1]{\begin{pmatrix}#1\end{pmatrix}}
\newcommand{\up}{\uparrow}
\newcommand{\down}{\downarrow}
\def\omE{\omega_{\rm Ter}}

%
%%%%%%%%%%     mauri    %%%%%%%%%%%%%%%%%%%%%%%%%%%%%%%%%

\newcommand{\Dsusy}{{susy \hspace{-9.4pt} \slash}\;}
\newcommand{\DCP}{{CP \hspace{-7.4pt} \slash}\;}
\newcommand{\mc}{\mathcal}
\newcommand{\gr}{\mathbf}
\renewcommand{\to}{\rightarrow}
\newcommand{\gtc}{\mathfrak}
\newcommand{\wh}{\widehat}
\newcommand{\br}{\langle}
\newcommand{\kt}{\rangle}

%%%%%%%%%%%%%%%%%%%%%%%%%%%%%%%%%%%%%%%%%%%%%%%%%%%%%%%%%%

% barbara Ricci  %definizione di minore e maggiore simile
\def\lsim{\mathrel{\mathop  {\hbox{\lower0.5ex\hbox{$\sim$}
\kern-0.8em\lower-0.7ex\hbox{$<$}}}}}
\def\gsim{\mathrel{\mathop  {\hbox{\lower0.5ex\hbox{$\sim$}
\kern-0.8em\lower-0.7ex\hbox{$>$}}}}}
%%%%%%%%%%%%%%%%%%%%%%%%%%%%%%%%%%

\def\nn{\\  \nonumber}
\def\de{\partial}
\def\brf{{\mathbf f}}
\def\bbf{\bar{\bf f}}
\def\bF{{\bf F}}
\def\bbF{\bar{\bf F}}
\def\bA{{\mathbf A}}
\def\bB{{\mathbf B}}
\def\bG{{\mathbf G}}
\def\bI{{\mathbf I}}
\def\bM{{\mathbf M}}
\def\bY{{\mathbf Y}}
\def\bX{{\mathbf X}}
\def\bS{{\mathbf S}}
\def\bb{{\mathbf b}}
\def\bh{{\mathbf h}}
\def\bg{{\mathbf g}}
\def\bla{{\mathbf \la}}
\def\bmu{\mathbf m }
\def\by{{\mathbf y}}
\def\bmu{\mbox{\boldmath $\mu$} }
\def\bsig{\mbox{\boldmath $\sigma$} }
\def\bunity{{\mathbf 1}}
\def\cA{{\cal A}}
\def\cB{{\cal B}}
\def\cC{{\cal C}}
\def\cD{{\cal D}}
\def\cF{{\cal F}}
\def\cG{{\cal G}}
\def\cH{{\cal H}}
\def\cI{{\cal I}}
\def\cL{{\cal L}}
\def\cN{{\cal N}}
\def\cM{{\cal M}}
\def\cO{{\cal O}}
\def\cR{{\cal R}}
\def\cS{{\cal S}}
\def\cT{{\cal T}}
\def\eV{{\rm eV}}

%
%%%%%%%%%%%%%%%%%%%%%%%%%%%%%%%%%%%%%

\title{Weak Gravity Conjecture as a Razor Criterium for Exotic D-brane instantons}

\author{Andrea Addazi$^1$}\email{andrea.addazi@infn.lngs.it}
\affiliation{$^1$ Dipartimento di Fisica,
 Universit\`a di L'Aquila, 67010 Coppito AQ and
LNGS, Laboratori Nazionali del Gran Sasso, 67010 Assergi AQ, Italy}

\begin{abstract}

We discuss implications of Weak gravity conjecture (WGC) for
exotic D-brane instantons. In particular, 
WGC leads to indirect stringent bounds on non-perturbative superpotentials 
generated by exotic instantons, with many implications  
for phenomenology: R-parity violating processes, neutrino mass,
$\mu$-problem, Neutron-Antineutron transitions and 
collider physics.

\end{abstract}

\maketitle
\section{Introduction}

The Weak Gravity Conjecture (WGC) states that the weakest force is gravity \cite{ArkaniHamed:2006dz}
\footnote{We mention that recently many papers on 
weak gravity conjecture 
in context of N-field and axion monodromy inflation appeared
 \cite{Ibanez:2015fcv, Brown:2015iha, Hebecker:2015rya, Hebecker:2015zss, Heidenreich:2015wga, Brown:2015iha, Brown:2015lia, Kooner:2015rza, Rudelius:2015xta, Junghans:2015hba, Brown:2015lia}. Other recent applications can be found in Refs.   
\cite{Cheung:2014ega,Cheung:2014vva,Li:2006vc,Heidenreich:2016jrl,Heidenreich:2016aqi,Montero:2016tif}.}. 
This implies that for each abelian gauge boson 
$U_{X}(1)$, there must be a light charged particle 
with mass $m$ and charge $q_{X}$
satisfying the bound 
\be{WHC}
\frac{m}{q_{X}}\leq M_{Pl}
\ee 
In other words, for every $U_{X}(1)$ gauge in 4-d
there is a new UV scale satisfying the bound
\be{bound}
\Lambda_{X}= g_{X}M_{Pl}= \frac{g_{X}}{\sqrt{G_{N}}}
\ee
The WGC is motivated by 
two main arguments: 
i) global abelian symmetries cannot exist in string theory;
ii) the absence of black hole remnants, motivated by holography
\cite{Susskind:1995da}.
 We mention that a possible test of the WGC applied to a long-range $U(1)_{B-L}$ 
can be in Neutron-Antineutron oscillation experiments
(See \cite{Addazi:2015pia,Addazi:2016rgo,Babu:2016rwa} for recent analysis of stringent limits on B-L couplings 
to neutrons (antineutrons)).

A more surprising result is that, according to Weak gravity conjecture, Eq.(\ref{WHC}) should be also true for 
magnetic monopoles. 
For example, for a t'Hooft-Polyakov monopole 
obtained by the Higgsing $SU(2)\rightarrow U(1)$, 
the Weak gravity conjecture sets a bound which reads
\be{monopol}
M_{mon}\leq g_{m}M_{Pl}\sim \frac{1}{g_{X}}M_{Pl}
\ee
where $g_{m}$ is the monopole magnetic charge. 
This bound is also understood by the fact that 
a fundamental monopole
cannot have an energy density higher than black hole one.  
As is well known, monopoles and gauge instantons are
always in a correspondence. 
In particular, gauge instantons in a d-dimensional YM 
theory correspond to monopole solutions in higher dimensional
YM theories closed by domain walls \cite{Dvali:2007nm}. 
In particular, the action of instantons is
$S=M_{mon}/v_{X}$, i.e. it is bounded by Eq.(\ref{monopol}). 
These arguments can be generalized in string theories.
In particular Euclidean D-brane instantons 
in IIA and IIB superstring theories 
can be understood as a generalization of the 
soliton/instanton correspondence:
a stringy instanton in the effective low energy 4-d theory
corresponds to a Euclidean D-brane in the 10-d superstring theory. 
 As is known, Euclidean D-(p{-}4) branes wrapping the same
n-cycles of ordinary D-p branes on the Calabi-Yau internal space
reproduce YM gauge instantons in the low energy 4-d gauge theory
$(\alpha'=l_{s}^{2}\rightarrow 0)$.  
However, it exists a new class of instantons,
dub {\it exotic instantons}, 
associated  to
Euclidean D-(p-4) branes wrapping different n-cycles 
than background D-p branes.
This implies that exotic instantons are not related 
to any gauge couplings of the $d=4$ effective gauge theory. 
Intriguingly, they can generate new effective superpotential
terms among fields even if not allowed at perturbative level
\footnote{D-brane instantons were studied in various quiver theories
without and with fluxes in Refs.\cite{Bianchi:2007wy,Billo:2008sp,Bianchi:2007rb,Bianchi:2012pn,Bianchi:2012ud}
(See also Ref.   \cite{Bianchi:2009ij} for a review on D-brane instantons).}. 
Recently, we discussed possible implications 
of exotic instantons in $B-L$ physics. 
In fact, contrary to gauge instantons, 
exotic instantons could violate $U(1)$ vector-like symmetries 
like $B-L$ with a strong coupling. 
For example, $\Delta B=2$ violating processes
like Neutron-Antineutron transitions can be 
generated from several different models, 
without destabilizing the proton 
\cite{Addazi:2014ila,Addazi:2015ata,Addazi:2015rwa,Addazi:2015hka,Addazi:2015eca,Addazi:2015fua,Addazi:2015oba,Addazi:2015goa,Addazi:2015ewa,Addazi:2015yna,Addazi:2016xuh,Addazi:2016mtn}. So that, it is important to identify 
a possible {\it razor principle}
for exotic instantons which, 
rephrasing Arkani-Hamed {\it et al} (in Ref. \cite{ArkaniHamed:2006dz}), 
could select {\it D-brane instantons of 
the Landscape against ones of Swamplands}. 
In this paper, we suggest an extension of the Weak gravity conjecture
to exotic stringy instantons. 
In particular, we will see how the weak gravity conjecture 
on exotic instantons can provide stringent constrains
for internal spaces and n-cycle geometries.
We will show how the WGC on exotic instantons leads
to interesting predictions in phenomenology.

\section{WGC bounds to E2-branes in IIA string theory }

In this section, we will discuss the WGC on Exotic Instantons 
in the specific context of IIA open string theories
with intersecting D6-brane stacks wrapping 3-cycles in the $CY_{3}$
and Euclidean D2-brane (E2-brane, exotic) instantons wrapping (different) 3-cycles in the $CY_{3}$.
Of course, the following considerations can be easily generalized to
other open string theories:
i) type IIB open string theories, with 
space-filling D3-branes and D7-branes 
wrapping holomorphic divisors in a $CY_{3}$
and (exotic) E3-branes wrapping (different) holomorphic divisors, 
ii) type I open string theories with magnetized D9-branes wrapping 
a $CY_{3}$
and (exotic) magnetized E5-branes

\begin{figure}[htb] \label{MDA}
\begin{center}
\caption{a) The intersections among the $D6$-brane stack (double line) and the $E2$-brane. Modulini 
$\tau,\alpha$ are localized among the $E2-D6$ intersections while the physical field is 
in the intersection $D6-D6$. b) The dual mixed disk amplitude with modulini $\alpha,\tau$ and the physical field $\Phi$.
In the limit $\alpha'\rightarrow 0$, the mixed disk amplitude generates an effective lagrangian term 
among $\alpha,\tau,\Phi$. Integrating on the modulini space, this leads to the generation of a new superpotential
dependent by the number of intersections among $D6$-brane stacks and $E2$-brane. 
   }
\includegraphics[scale=0.06]{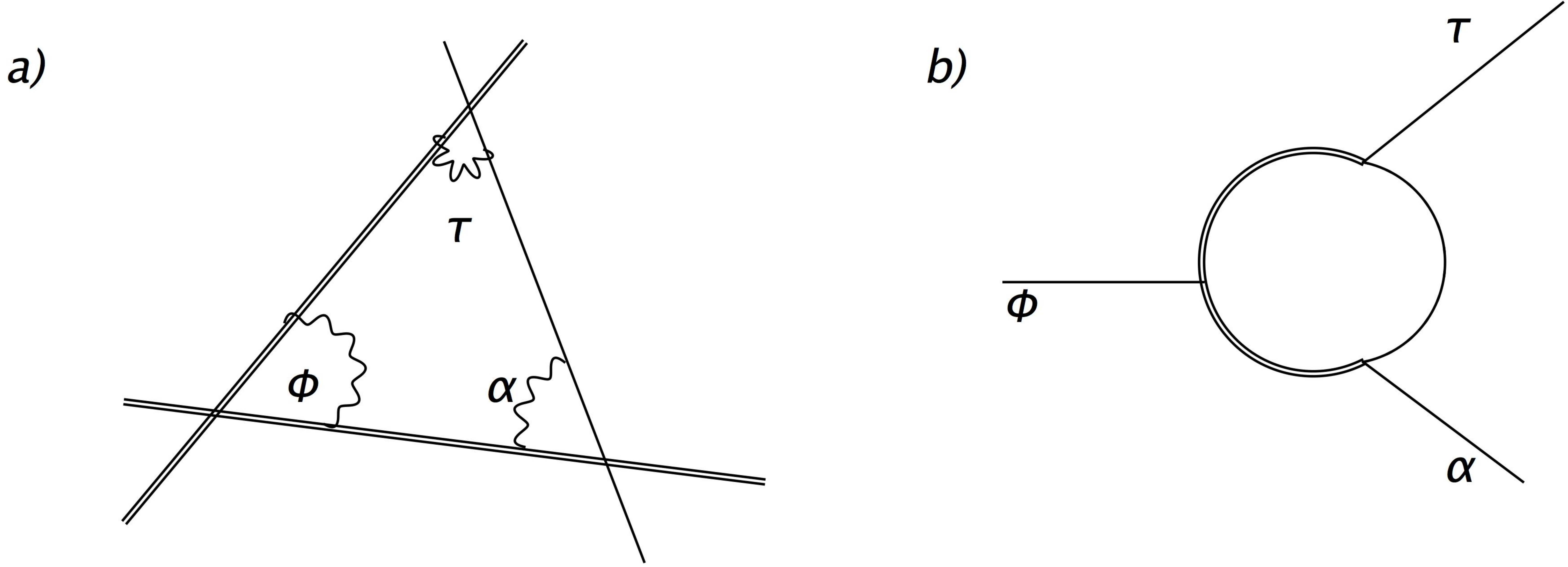}  
\end{center}
\vspace{-1mm}
\end{figure}

\subsection{Theoretical preliminaries}

In type IIA open string theories, 
E2-branes can intersect D6-brane stacks
and open strings localized among these intersections 
correspond to fermionic zero modes of the exotic instantons. 
In the low energy limit $\alpha'\rightarrow 0$, 
\be{generation}
\mathcal{W}_{n.p.}=e^{-S_{E2}} M_{S}^{-N+3}\Phi^{N}
\ee
where $\Phi^{N}$ is a charged combination 
of a $U(1)$ gauge symmetry. 
The microscopic mechanism generating Eq.(\ref{generation})
is displayed in Fig.1 (see Ref.\cite{Bianchi:2009ij} for a complete review on these aspects). 
The imaginary part of the instanton
action is related to the pseudoscalar
field shifting under  a $U(1)$ gauge symmetry.
The relevant $U(1)$ is broken 
by a St\"ckelberg mechanism 
due to a coupling of its gauge curvature 
with closed strings (RR fields)
as $B\wedge F$. 
The instanton coupling contains the instanton action 
which is complex and given by DBI+CS terms:
\be{S}
Re \, S_{E2}=\mathcal{V}_{\Pi_{3}}g_{S}^{-1},\,\,\,Im \,S_{E2}=\sum_{r}c_{r}a_{r}
\ee
where $a_{r}$ are the RR axions which are coupled to the $E2$-brane. 

For a $D6$-brane models, 
we generically have a set of different
D6-brane stacks labeled as $A,B,C,...$.
The 3-cycles $\Pi_{A}$ of the $D6_{A}$-brane 
can be decomposed in a basis $\{C_{r}\}$
of homology 3-cycles: 
 \be{Cycles}
 \Pi_{A}=\sum_{r}p_{Ar}C_{r}
 \ee
On the other hand, the coupling of a second 
$D6_{B}$-brane to the RR axions 
$a_{r}$ demands 
the shift symmetry 
\be{shift}
a_{r}\rightarrow N_{B}(p_{Br}-p_{B^{*}r})\Lambda_{B}
\ee
This implies the shift symmetry
\be{SE2}
e^{-S_{E2}}\rightarrow e^{-i\sum_{A}N_{A}(I_{MA}-I_{MA^{*}})\Lambda_{A} }e^{-S_{E2}}
\ee
Constrains in the intersection numbers of $E2$-branes and $D6$-branes
3-cycles:
\be{cycles}
I_{Ma}-I_{Ma^{*}}=I_{Mb}-I_{Mb^{*}}=0
\ee
\be{cycles2}
I_{Mc}-I_{Mc^{*}}=I_{Md}-I_{Md^{*}}=2
\ee
where $I_{Ma}=\Pi_{M}\cdot \Pi_{a}$ are the numbers of intersections
among the $E2$-brane and $D6$-branes. 
Under this condition, the shift term of $\Phi^{N}$ 
under $U(1)$ is exactly compensated by 
the shift symmetry of the instanton coupling. 

\subsection{ Weak gravity conjecture and E2-branes}

The E2-brane is charged with respect to $n$ RR fields
in turns related to the St\"ueckelberg breaking of $n$ 
(anomalous and non-anomalous) $U(1)$s.
On the other hand, 
ordinary E2-branes corresponding to gauge instantons are
constrained by the low energy limit bound
(\ref{monopol}), $M\leq g_{YM}^{-1}M_{Pl}$
where $g_{YM}^{2}=(2\pi)^{4}g_{S}\mathcal{V}_{\Pi_{3}}^{-1}$.
The generalized WGC bound on a Dp-brane with tension $T_{p}$ and  sum of Dp-brane charges $Q$ is 
\be{MassBound}
\frac{T_{p}}{Q}\leq M_{Pl}^{p+1}
\ee
For an E2-brane, Eq.(\ref{MassBound}) corresponds to 
\be{GWGC1}
\frac{\mathcal{V}_{\Pi_{3}}}{g_{s}}\leq \left(\sum_{r}c_{r}\right)\left( \frac{M_{Pl}}{M_{s}}\right)^{3}=\left(\sum_{r}c_{r}\right)\left(\frac{\sqrt{\mathcal{V}_{6}}}{(2\pi)^{7/2}g_{s}}\right)^{3}
\ee
which in turn corresponds to 
\be{WGC1a}
|e^{S_{E2}}|\leq e^{\sqrt{2}\left(\sum_{r}c_{r}\right)\left(\frac{\sqrt{\mathcal{V}_{6}}}{(2\pi)^{7/2}g_{s}}\right)^{3}}
\ee

%Second bound:
%\be{GWGC2}
%e^{+S_{E2}}\leq \left(\frac{M_{Pl}}{M_{s}}\right)^{N-3}=\left(\frac{\sqrt{\mathcal{V}_{6}}}{(2\pi)^{7/2}g_{s}} \right)^{N-3}
%\ee

%\be{GWGC}
%M\leq M_{Pl}\sqrt{\left(\sum_{r=1}^{n}c_{r}\right)^{2}+\frac{\mathcal{V}_{\Pi_{3}}}{(2\pi)^{4}g_{S}}}
%\ee
%where $c_{r}$ are the RR-charges of the E2-brane.
%On the other hand, $M^{N-3}=e^{+S_{E2}}M_{S}^{N-3}$. 
%As a consequence, Eq.(\ref{GWGC}) sets 
%a bound on the instantonic action that reads
%\be{SE2}
%|e^{+S_{E2}}|\leq \left\{\frac{\mathcal{V}_{6}}{(2\pi)^{7}g_{s}^{2}}\left[\left(\sum_{r=1}^{n}c_{r}\right)^{2}+\frac{\mathcal{V}_{\Pi_{3}}}{(2\pi)^{4}g_{S}}\right]\right\}^{(N-3)/2}
%\ee
%where $M_{Pl}^{2}=M_{S}^{2}\mathcal{V}_{6}/((2\pi)^{7}g_{s}^{2})$. 
%On the other hand, 
%ordinary E2-branes corresponding to gauge instantons are
%constrained by the low energy limit bound
%(\ref{monopol}), $M\leq g_{YM}^{-1}M_{Pl}$
%where $g_{YM}^{2}=(2\pi)^{4}g_{S}\mathcal{V}_{\Pi_{3}}^{-1}$:
%\be{New}
%M\leq \frac{1}{4\pi^{2}\sqrt{g_{S}}}\mathcal{V}_{\Pi_{3}}^{1/2}M_{Pl}
%\ee
%We conjecture that Eq.(\ref{New}) can be generalized to Exotic instantons, 
%where the combination $(2\pi)^{4}g_{S}\mathcal{V}_{\Pi_{3}}^{-1}$ is not related 
%to any gauge coupling \footnote{$\mathcal{V}_{6}$ and $\mathcal{V}_{\Pi_{3}}$ 
%are adimensional in units of $l_{s}=1$.}.
%Eq.(\ref{New}) implies the following bound on the instantonic coupling: 
%\be{Boundf}
%|e^{+S_{E2}}|\leq \left(\frac{1}{(2\pi)^{11/2}}\frac{\sqrt{\mathcal{V}_{\Pi_{3}}\mathcal{V}_{6}}}{g_{s}^{3/2}}\right)^{N-3}
%\ee

The violation of Eq.(\ref{GWGC1}) implies that an extremal black hole remnant with 
RR E-brane charges equal to the black hole mass (normalized in Planck units)
could be formed, which is in contrast with holographic arguments. 

%The violation of Eq.(\ref{GWGC2}) would imply that a N-particles interaction 
%suppressed as $M^{N-3}$ and mediated by an E2-brane should be weaker than a gravitational interaction among the same N-particles,
%which is tree-level suppressed as $G_{N}^{N-3}\sim M_{Pl}^{-2N+6}$. 
%However, supposing for example $(M/M_{Pl})<10^{-1000}$,
%this implies that a black hole can have 
%a charge among $M/M_{Pl}=0,10^{1000}/(N-3)$.
%Group by group of 

%As a consequence, an extremal BH remnant with N'\times (N-3)$ particles with charge 
%$Q=N' 

%$N'M=N'M_{Pl}$  
%would saturate 
%

The WGC bounds should be combined with 
the semiclassical approximation bound
\be{WGC2a}
|e^{S_{E2}}| \geq e^{\sqrt{\left(\frac{\mathcal{V}_{s}}{g_{s}}\right)^{2}+\left(\sum_{r}c_{r}\right)^{2}}}
\ee
where $\mathcal{V}_{s}$ is the string volume 
of the 3-cycles. For example, for a 3-cycle cilindrical geometry, 
$\mathcal{V}_{s}=\pi^{2}$.  
Essentially, Eq.\ref{bound} is an universal stringy bound on the 
instanton-antinstanton bubbles or instanton partition functions. 

So that, weak gravity conjecture constrains 
3-cycle size wrapped by the E2-branes, 
the string coupling the internal volume 
(the string scale over the planck scale). 

Finally, if $M\geq M_{Pl}$
in the superpotential 
Ref.(\ref{generation}) for $N>3$, 
quantum gravity (mainly corrections to the superpotential from reggeized gravitons exchanges) corrections should affect
our instantonic calculations. 
So that, we impose a calculability bound 
\be{WGC3a}
|e^{S_{E2}}|\leq \left(\frac{M_{Pl}}{M_{s}}\right)^{N-3}=\left(\frac{\sqrt{\mathcal{V}_{6}}}{(2\pi)^{7/2}g_{s}}\right)^{N-3}
\ee
for $N>3$. 
On the other hand, for $N=2$ (bilinear superpotentials), $M\leq M_{Pl}$
in order to neglect quantum gravity corrections
and 
\be{WGC3b}
|e^{-S_{E2}}|\leq \left(\frac{M_{s}}{M_{Pl}}\right)=\left(\frac{\sqrt{\mathcal{V}_{6}}}{(2\pi)^{7/2}g_{s}}\right)^{-1}
\ee

%\begin{figure}[t]
%\centerline{ \includegraphics [height=8cm,width=0.7\columnwidth]{...}}
%\vspace*{-1ex}
%\caption{a)  }
%\label{plot}   % \ref{plot}
%\end{figure}
 
 \section{Implications in phenomenology}
 
 \begin{figure}[htb] \label{WGC2}
\begin{center}
\caption{(Un)orientied quiver theory embedding the (MS)SM and generating through an Exotic E-brane Instanton a six quark superpotential 
(See Eq.(\ref{NNBAR}). This model was discussed in companion papers \cite{Addazi:2015goa,Addazi:2015ewa}.
For instance, this quiver is (locally) compatible with 
the tadpole cancellations and SM hypercharge conditions 
discussed and applied in Refs. \cite{Cvetic:2009yh,Cvetic:2010dz,Cvetic:2011iq,Cvetic:2015vit}.}
\includegraphics[scale=0.09]{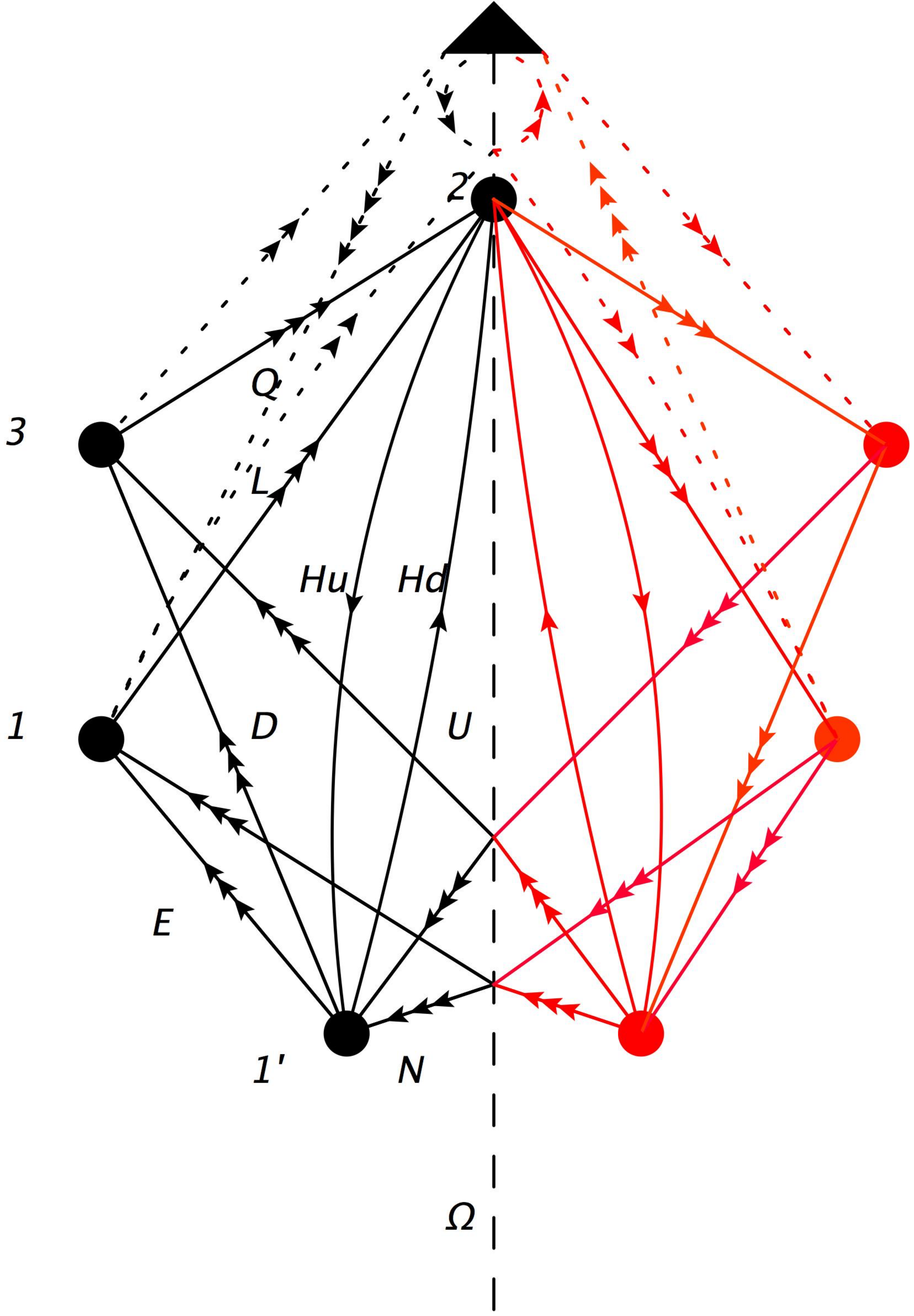}  
\end{center}
\vspace{-1mm}
\end{figure}

\begin{figure}[htb] \label{WGC3}
\begin{center}
\caption{In the same model of Fig.2, an Exotic E-brane instanton generating a Weinberg operator 
as Eq.(\ref{W}) is displayed.  }
\includegraphics[scale=0.08]{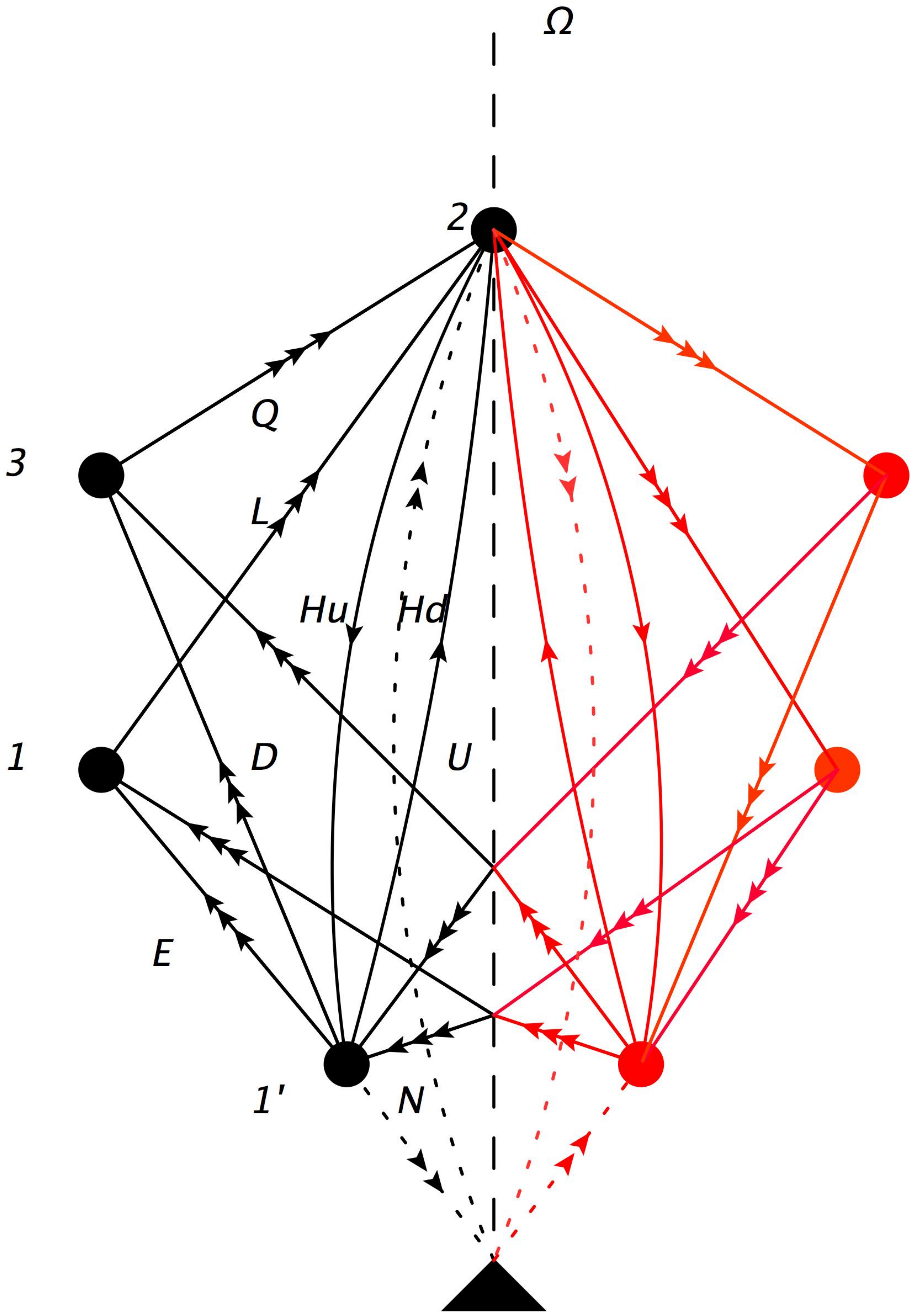}  
\end{center}
\vspace{-1mm}
\end{figure}

In this section, we will discuss the application of WGC 
and the semiclassical bound for
a set of superpotentials 
generated by exotic E-brane instantons
studied in literature 
\cite{Ibanez:2006da,Blumenhagen:2006xt,Ibanez:2007rs,Blumenhagen:2009qh,Blumenhagen:2007bn,Blumenhagen:2007zk,Abe:2015uma,Addazi:2014ila,Addazi:2015ata,Addazi:2015rwa,Addazi:2015hka,Addazi:2015eca,Addazi:2015fua,Addazi:2015oba,Addazi:2015goa,Addazi:2015ewa,Addazi:2015yna,Addazi:2016xuh,Addazi:2016mtn}.

For Trilinear Yukawa couplings, the window allowed by WGC and semiclassical bounds is
\be{SE2}
e^{-\sqrt{2}\left(\sum_{r}c_{r}\right)\left(\frac{\sqrt{\mathcal{V}_{6}}}{(2\pi)^{7/2}g_{s}}\right)^{3}} \leq |e^{-S_{E2}}| \leq e^{-\sqrt{\left(\frac{\mathcal{V}_{s}}{g_{s}}\right)^{2}+\left(\sum_{r}c_{r}\right)^{2}}}
\ee
These constrains allow a various set of possible coupling $e^{|S_{E2}|}$. 
For example, for high scale string theory 
$M_{s}\simeq 10^{15}\, \rm GeV$, $\mathcal{V}_{6}g_{s}\simeq 1$,
$g_{s}\simeq 10^{-3}$, WGC lower bound is easily satisfied
($|e^{-S_{E2}}|\geq e^{-10^{9}}$ or so), 
while the semiclassical upper bound is expected to be of order one
because of RR charge contribution. 
So that, R-parity violating operators
destabilizing the proton like 
\be{neutrino}
Y_{1}e^{-S_{E2'}}U^{c}D^{c}D^{c},\,\,\,Y_{2}e^{-S_{E2''}}LQD^{c},\,\,\,Y_{3}e^{-S_{E3'''}}LLE^{c}
\ee
can be naturally suppressed without violating the WGC. 
As well as  the generation of a small Dirac mass term for the neutrino like
\be{unnatural}
\mathcal{L}=y_{\nu}e^{-S_{E_{2}}} \langle H \rangle \bar{\nu}_{L}\nu_{R}
\ee
without any see-saw mechanisms. 
However, we should note that this imposes a bound on the number of RR charges.
As a consequence, a quiver theory with a large number of D-brane stacks (order $N_{stacks}>10^{3}$)
would enhance the WGC bound for non-perturbative trilinear couplings. 
This seems to imply that SM has to be embedded in a quiver with few D-brane stacks nodes
(this is also desired for minimality). 

For bilinear terms like the $\mu$-term
and $\mu' H L$, with 
$\mu=M_{S}e^{-S_{E2}}$ 
and similarly $\mu'$, 
a combination of Eqs.(\ref{WGC1a},\ref{WGC2a},\ref{WGC3b}).

This means that a scenario in which $M_{S}\simeq M_{GUT}\simeq 10^{15}\, \rm GeV$
with $\mu\simeq TeV$ $(|e^{-S_{2}}|\simeq 10^{-12}$) 
is constrained by WGC, because $\mathcal{V}_{6},\mathcal{V}_{\Pi_{3}}\geq 1$ while $g_{s}\leq 1$. 
In particular, 
$$e^{-\sqrt{2}\left(\sum_{r}c_{r}\right)\left(\frac{\sqrt{\mathcal{V}_{6}}}{(2\pi)^{7/2}g_{s}}\right)^{3}}\leq |e^{-S_{E2}}|\leq 10^{-12}$$
implying  
$$\sqrt{2}\left(\sum_{r}c_{r}\right)\left(\frac{\sqrt{\mathcal{V}_{6}}}{(2\pi)^{7/2}g_{s}}\right)^{3}>12$$
which can be easily satisfied in perturbative string theory. 

Let us now discuss a proton decay superpotential 
generated by E2-brane instantons:
\be{qqql}
\mathcal{W}=\frac{e^{-S_{E2}}}{M_{s}}Q Q Q L\rightarrow \mathcal{O}_{p{-}decay}=\frac{e^{-S_{E2}}}{M_{s}M_{SUSY}}qqql
\ee
which is constrained up to 
$$M_{s}M_{SUSY}e^{+S_{E2}}>(10^{15}\, \rm GeV)^{2}$$
Let us suppose a scenario with $M_{s},M_{SUSY}\simeq 1\div 100\, \rm TeV$, 
In this case, the $e^{+S_{E2}}>10^{18}\div 10^{24}$, 
which constrains $(g_{s}^{-1},\mathcal{V}_{\Pi_{3}},\mathcal{V}_{6})>>1$.
In the opposite regime $M_{S}\simeq M_{SUSY}\simeq 10^{15}\, \rm GeV$,
the coupling $e^{+S_{E2}}$ could be much smaller in order to satisfy the 
experimental constrains, 
i.e. $\mathcal{V}_{6},\mathcal{V}_{3}\simeq 1$.
However,  semiclassical and quantum gravity bounds imply that the effective proton-decay scale is 
$M_{s}\leq M_{p-decay}\leq M_{Pl}$. The same proton-decay bounds can be applied to 
E2-branes (directly) generating a Weinberg superpotential
as
\be{W}
\mathcal{W}=\frac{e^{-S_{E2}}}{M_{s}}H_{u}LH_{u}L\rightarrow \frac{e^{-S_{E2}}}{M_{s}M_{SUSY}}hlhl
\ee
Assuming $M_{s},M_{SUSY}\simeq 1\div 100\, \rm TeV$, the effective scale
$e^{+S_{E2}}M_{s}M_{SUSY}\simeq 10^{12}\div 10^{13}\, \rm GeV$.
So that, in analogy with proton-decay operator, $e^{+S_{E2}}>>1$ ($e^{+S_{E2}}\simeq 10^{15}\div 10^{21}$). 
As for proton decay, WGC demands $(g_{s}^{-1},\mathcal{V}_{\Pi_{3}},\mathcal{V}_{6})>>1$.
A scenario with $M_{SUSY}\simeq M_{S}\simeq 10^{15}\, \rm GeV$
is not compatible for any value of $e^{+S_{E2}}$ 
with a scale $10^{12}\div 10^{13}\, \rm GeV$ for the neutrino Majorana mass. 
A possible scenario which is compatible with WGC is 
$e^{+S_{E2}}\simeq 1$ while $M_{s}M_{SUSY}\simeq (10^{12}\div 10^{13}\, \rm GeV)^{2}$. 
However, $e^{+S_{E2}}\simeq 1$ violates the semiclassical bound. 

In Refs.\cite{Addazi:2015goa,Addazi:2015ewa}, we suggested that
a six quark $\Delta B=2$ transition can be generated by only one exotic instanton solution
as
\be{NNBAR}
\mathcal{W}=\frac{1}{M_{S}^{3}}e^{-S_{E2}}\mathcal{Y}_{f_{1}f_{2}f_{3}f_{4}f_{5}f_{6}}U_{f_{1}}^{c}D_{f_{2}}^{c}D_{f_{3}}^{c}U_{f_{4}}^{c}D_{f_{5}}^{c}D_{f_{6}}^{c}
\ee
After the supersymmetric quark-squark reduction (mediated by gauginos)
, this operator can generate an effective neutron-antineutron operator 
\be{Ope}
\mathcal{O}_{n\bar{n}}=\frac{Y_{1}}{\Lambda_{n\bar{n}}^{5}}u^{c}d^{c}d^{c} u^{c}d^{c}d^{c}
\ee
$$\Lambda_{n\bar{n}}^{5}=e^{+S_{E2}}M_{S}^{3}M_{SUSY}^{2}\geq (300\, \rm TeV)^{5}$$
where $300\, \rm TeV$ is the current best experimental bound. 
Next generation of experiments can enhance the current bound up to $1000\, \rm TeV$ scale. 
Let us assume that $M_{SUSY}\simeq M_{S}$.
In principle, in order to generate a  $1000\, \rm TeV$ $n-\bar{n}$ transition with a $M_{S}=10\, \rm TeV$, 
one could consider a $Y_{1}|e^{S_{E2}}|\simeq 10^{10}$, which corresponds to large 3-cycles wrapped by the $E2$.
However, the weak gravity conjecture sets a bound on the largeness of the (inverse) instanton coupling. 
For example, in a scenario in which $g_{s}^{-2}\mathcal{V}_{6}\simeq 1$, 
a large instantonic mass scale is excluded: the string scale must be $M_{s}<<M_{Pl}$, i.e.
 $\mathcal{V}_{6}>> g_{s}^{2}$. 
So that, in this scenario Regge stringy states 
can be searched in the next run of LHC
while KK-modes cannot be TeV-ish particles.

A situation with $M_{S}\simeq M_{SUSY}\simeq 1000\, \rm TeV$, $Y_{1}e^{+S_{E2}}\simeq 1$ is possible 
but cannot be controlled in semiclassical instantonic methods. Such a situation corresponds 
to collapsed three-cycles on a singularity in the $CY_{3}$ ($\mathcal{V}_{\Pi_{3}}<<1$). 
This case can be easily compatible with WGC in a larger landscape of parameters
with respect to the case discussed above.
However, this scenario is less appealing from the calculation point of view (non-perturbative uncontrolled regime). 

A scenario which is interesting for future $100\, \rm TeV$ proton-proton colliders beyond LHC
is
$M_{S}\simeq M_{SUSY}\simeq 50 \div 100\, \rm TeV$.
In this case, a $1000\, \rm TeV$ neutron-antineutron transition can be generated 
for
$Y_{1}|e^{+S_{E2}}|\simeq 10^{5}\div 10^{6}$, which is not viable for $g_{s}^{-2}\mathcal{V}_{6}\simeq 1$
$(M_{S}\simeq M_{Pl})$ because of WGC. So that, in this scenario $M_{s}<<M_{Pl}$, i.e.
Regge states and no KK modes  for a $100\, \rm TeV$ proton-proton collider.
In this case, there is also the interesting possibility to observe the $\Delta B=2$ exotic instanton 
in direct binary quark collisions, as $qq\rightarrow \tilde{q}\tilde{q}\tilde{q}\bar{\tilde{q}}$.
In fact, while neutron-antineutron transitions are constrained up to $300\, \rm TeV$, other six quarks 
operators involving heavier flavors can be tested in $100\, \rm TeV$ colliders,
i.e. the Yukawa matrix components of other flavors 
are not directly constrained by $n-\bar{n}$ experiments. 
For instance, the cross section is expected to be polynomially growing with the CM energy 
for $E_{CM}<M$, while unitarized and exponentially softened for $E_{CM}>>M$ at fixed scattering angle
\cite{Addazi:2015ewa}. 
  
\section{Conclusions and remarks}  
  
In this letter, we have discussed many implications of the Weak Gravity Conjecture for
exotic stringy instantons. In particular, we have considered the case of type IIA
open string theory. 
We have shown how WGC on exotic instantons 
can provide a strong bound on effective superpotentials generated by E-branes. 
For instance,  WGC constrains 
 3-cycles geometries of the E2-brane instantons, the string coupling 
and the internal six-dimensional volume. 
For example, we have argued that, if a $1000\, \rm TeV$ neutron-antineutron transition 
was observed, WGC, applied on the direct generation mechanism from a $E2$-brane instanton
in low scale string theory,
would imply a precise set of observables for the next generation of colliders. 
Assuming the string scale $M_{s}=10\div 100\, \rm TeV$, we have shown 
how  the internal space volume must be very large, i.e. 
$100\, \rm TeV$ proton-proton colliders beyond LHC should observe Regge string states and not KK states. 

We have also discussed WGC implications on bilinear and trilinear R-parity violating operators, 
Dirac neutrino mass, proton decay operators and $\mu$-term from exotic instantons. 

We conclude that the Weak Gravity Conjecture, 
which is a statement on quantum gravity and black holes, 
seems to 
be unexpectedly predictive in string-inspired particle physics.
Exotic instantons may have an important 
role in the UV completion of the standard model 
of particles and cosmology from  basic principles of the string theory. 
WGC could be a razor criterium which may be 
 crucially important 
for our understanding of string phenomenology.

\vspace{0.1cm}

{\large \bf Acknowledgments}
\vspace{3mm}

I would like to thank organizers and 
participants of String Phenomenology 2016 (Ioannina, Greece)
for inspiring talks and conversations, 
as well as A.P. Wine for interesting suggestions and remarks.
My work was supported in part by the MIUR research grant Theoretical Astroparticle Physics PRIN 2012CP-PYP7 and by SdC Progetto speciale Multiasse La Societ\'a della Conoscenza in Abruzzo PO FSE Abruzzo 2007-2013.

\end{document}